\documentclass[aps,pra,twocolumn,floatfix]{revtex4}
\usepackage{eurosym}
\usepackage{amsfonts}
\usepackage{mathrsfs}
\usepackage{color}
\usepackage{xspace}
\usepackage{subfigure}
\usepackage{longtable}
\usepackage{xspace}
\usepackage{multirow}
\usepackage{tabularx}
\usepackage{amsmath}
\usepackage{amssymb}
\usepackage{graphicx}
\usepackage{dcolumn}
\usepackage{bm}
\usepackage[colorlinks,urlcolor=blue,citecolor=blue,linkcolor=blue]{hyperref}
\usepackage[normalem]{ulem}
\usepackage{epstopdf}
\usepackage{appendix}

\allowdisplaybreaks

\begin{document}

\title{Topological phases in pseudospin-1 Fermi gases with two-dimensional
spin-orbit coupling}
\author{Junpeng Hou}
\author{Haiping Hu}
\author{Chuanwei Zhang}
\thanks{chuanwei.zhang@utdallas.edu}
\affiliation{Department of Physics, The University of Texas at Dallas, Richardson, Texas
75080, USA}

\begin{abstract}
The recent experimental realization of spin-orbit (SO) coupling for
ultracold bosons and fermions opens an exciting avenue for engineering
quantum matter that may be challenging to realize in solid state materials
such as SO coupled pseudospin-1 fermions. While one-dimensional SO coupling for
spin-1 bosons has been experimentally realized, the generation of
two-dimensional (2D) SO coupling and its topological properties are largely
unexplored. Here we propose an experimental scheme for realizing a 2D
Rashba-type SO coupling in a square lattice for pseudospin-1 Fermi gases. Because
of the extended spin degree of freedom, many interesting topological phases
could exist without relying on lattice point group symmetries that are
crucial in solid state materials. These exotic phases include
triply-degenerate points, quadratic band touching, a large Chern number ($%
C=5 $) superfluid with 5 Majorana modes, triple-Weyl fermions, etc. Our
scheme can be generalized to larger spins and provides a new route for
engineering topological quantum matter by utilizing large spin degrees of
freedom, instead of specific lattice symmetries.
\end{abstract}

\maketitle

\section{Introduction}

Spin-orbit (SO) coupling, the interaction between spin and orbital (e.g.,
momentum) degrees of freedom of a particle, plays an important role in many
topological phases of matter. In ultracold atomic gases, synthetic SO
coupling has been realized by coupling atomic hyperfine ground states
(denoted as pseudospins, but sometimes abbreviated as spin if there is no ambiguity) using
Raman lasers that induce momentum changes between different spin states. In
particular, both 1D and 2D SO couplings have been realized in experiments
for pseudospin-1/2 bosons and fermions and their distinct properties have been
widely studied \cite{QuC2013,LinYJ2011,ZhangJY2012,WangP2012,CheukLW2012,
WilliamsRA2013,OlsonAJ2014,WuZ2016,HuangLH2016,MengZM2016,Sun2017,ValdesCuriel2019}.

Recently 1D SO coupling for spin-1 bosons has also been experimentally
realized \cite{CampbellDL2016,LuoX2016}, which hosts some interesting
quantum phases \cite%
{LanZ2014,NatuSS2015,SunK2016,YuZQ2016,MartoneGI2016,KonigEJ2018}. Different
from electron's spin-1/2, the large number of available hyperfine states
provide a platform for studying fermionic atoms with integer pseudospins such as
pseudospin-1, which generally are difficult to realize in solid state materials.
The existence of such extra spin states naturally posts two important
questions: Can important topological physics emerge from 2D SO coupled
pseudospin-1 Fermi gases? If so, how can pseudospin-1 2D SO coupling be realized in
realistic experimental systems?

In this paper, we address these two important questions by showing that
many exotic topological phases can emerge from a pseudospin-1 degenerate Fermi gas
in a square optical lattice with 2D Rashba-type SO coupling, which can be
realized with a simple laser setup. These topological phases originate from
the coupling with the extra spin state in spin-1, instead of certain lattice
symmetries that dictate many topological solid state materials. Our main
results are:

\textit{i}) In the absence of Zeeman field to lift the degeneracy of three
pseudospin states at the center of the Brillouin zone (BZ), there exists a single
triply degenerate point in the 2D single particle band structure, which
consists of two linear and one flat bands. Three-dimensional
triply-degenerate points have been theoretically proposed in solid state,
ultracold atomic and optical systems with some experimental evidences \cite%
{BradlynB2016,WinklerGW2016,WengHM2016,ZhuZM2016,WengHM20161,LvBQ2017,
MaJZ2018,HuH2018,HouJ2018}. The 2D triply-degenerate point here resembles a
Dirac point in graphene without the valley degree of freedom \cite{Xiao08}.
A spin-tensor Zeeman field breaks the triple degeneracy, leading to a
quadratic band touching point due to indirect second-order spin coupling.
Quadratic band touching points have attracted great attention recently due
to their non-linear dispersions \cite{SunK2008,SunK2009,HerbutIF2014} and
many-body interaction driven quantum anomalous Hall ground states with
time-reversal symmetry breaking~\cite%
{SunK2008,SunK2009,PujariS2016,Wu2016,Zhu2016b,Chen2017}.

\textit{ii}) In the presence of attractive \textit{s}-wave pairing
interaction, 2D superfluids can become topological with large Chern numbers
up to $\pm 5$. The topological phase transition between different phases can
be accompanied with the band gap closing at (up to 2) points with cubic band
touching through pairing and indirect spin coupling, yielding the largest
Chern number change 6. The large Chern number topological superfluid can
host up to 5 Majorana edge states \cite{WilczekF2009,AliceaJ2012,FranzM2013}
simultaneously at the boundary.

\textit{iii}) In a 3D superfluid with 2D SO coupling, each cubic band
touching point becomes two triple-Weyl nodes located at $\pm k_{z}$ due to
the change of the effective chemical potential through the kinematic energy $%
\sim k_{z}^{2}$. Multi-Weyl fermions have attracted great attention due to
their multiple monopole charges and unusual transport properties in solid
state materials \cite%
{MaiXY2017,LiuQ2017,FangC2012,HuangZM2017,ZhangSX2016,HuangSM2016}.

\textit{iv}) An experimental setup for realizing 2D Rashba-type SO coupling
for pseudospin-1 atomic gases is proposed based on recently experimental success
for realizing 2D SO coupling for pseudospin-1/2 atoms \cite{WuZ2016}.

\textit{v}) Extending the findings to a higher pseudospin $\mathit{s}$, we show
that a high-order band touching point at the order of $2s$ and $4s-1$ can
exist for single-particle bands and pairing superfluids, respectively.

\section{Hamiltonian and single particle band topology} \label{sec_sp}

We consider a Rashba-type SO coupled pseudospin-1 Fermi gas confined in a square
lattice with both vector (linear) and tensor (quadratic) Zeeman fields. The
single particle Hamiltonian in the momentum space can be written as
\begin{equation}
H_{0}=-d_{x}F_{x}-d_{y}F_{y}+d_{z}(I-\frac{1}{2}F_{z}^{2}) +\frac{1}{2}%
\delta_{V}F_{z}+\frac{1}{4}\delta_{T}F_{z}^{2},  \label{ModelH}
\end{equation}%
under the three spin basis $\left\{ \left\vert 1\right\rangle ,\left\vert
0\right\rangle ,\left\vert -1\right\rangle \right\} $, where $d_{x}=2t_{%
\text{so}}\sin (k_{y})$, $d_{y}=2t_{\text{so}}\sin (k_{x})$, $d_{z}=2t(\cos
(k_{x})+\cos (k_{y}))$, $F_{i}$ represent the spin-$1$ vector operators. They can be expressed as $[F_x]_{mm'}=\sqrt{2}(\delta_{m-1,m'}+\delta_{m+1,m'})$, $[F_y]_{mm'}=\sqrt{2}i(\delta_{m-1,m'}-\delta_{m+1,m'})$ and $[F_z]_{mm'}=2m\delta_{m,m'}$ (see Appendix~\ref{appa}). $\delta _{T}$ and $\delta _{V}$ denote the tensor and
vector Zeeman fields. The experimental scheme for realizing this Hamiltonian
will be discussed later in the paper. Hereafter we take $t=t_{\text{SO}}=1$
for simplicity of the presentation.

\begin{figure}[t]
\centering\includegraphics[width=0.48\textwidth]{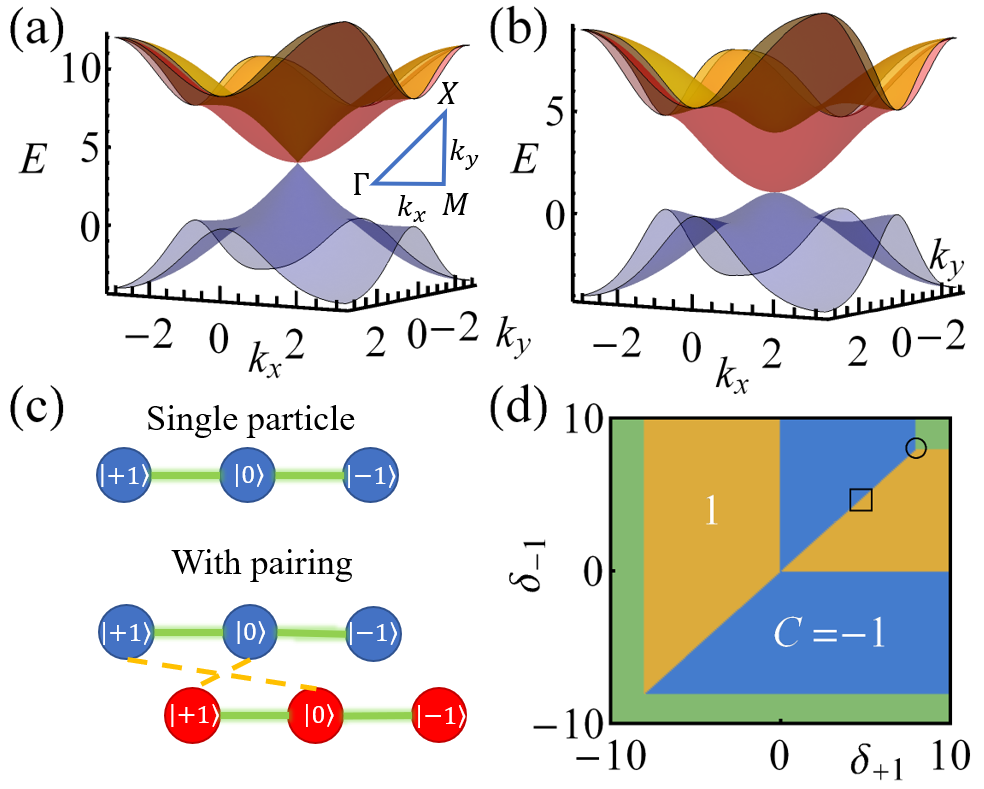}
\caption{(a) A 2D triply-degenerate point carrying $-2\protect\pi $ Berry
phase locates at $\Gamma $ point for $\protect\delta _{T}=8$ and $\protect%
\delta _{V}=0$. A quadratic band touching appears at X/M points for two
upper bands. Inset shows high-symmetry points in the BZ for a square
lattice. (b) Two quadratic band touchings locate at $\Gamma $ (between two
lower bands) and X/M (two upper bands) points for $\protect\delta _{T}=5$
and $\protect\delta _{V}=0$. (c) Coupling scheme with/without superfluid
pairing. The green line represents SO coupling, which contributes a winding
number $-1$. The dashed yellow line represents the \textit{s}-wave pairing,
which does not contribute any winding. The blue and red branches correspond
to particles and holes. (d) Phase diagram of Chern number for the lower
band. The circle and square denote the parameters for (a) and (b),
respectively.}
\label{fig1}
\end{figure}

When the vector Zeeman field $\delta _{V}=0$, the time-reversal symmetry of
the system is preserved. In this region, when the tensor Zeeman field $%
\delta _{T}=\pm 8$ or $0$, one (at $\Gamma $ or M point) or two (at X
points) 2D triply-degenerate points appear in the band structure, each of
which carries a topological charge (winding number) $-2$. An example of the
triply-degenerate point at $\Gamma $ point for $\delta _{T}=8$ is plotted in
Fig.~\ref{fig1}(a). The low-energy effective Hamiltonian around $\Gamma $ is
$\sim -\left( k_{y}F_{x}+k_{x}F_{y}\right) $ up to some constants, which can
be taken as a natural extension of the spin-1/2 Rashba SO coupling $%
k_{y}\sigma _{x}+k_{x}\sigma _{y}$ ($\sigma _{i},i=x,y,z$ are Pauli
matrices).

The triple degeneracy at $\Gamma $ for $\delta _{T}=8$ can be lifted by
varying $\delta _{T}$ and $\delta _{V}$. When the time-reversal symmetry is
still preserved (i.e., $\delta _{V}=0$), the decrease of $\delta _{T}$
lefts the top band, leaving a quadratic band touching between two bottom
bands (Fig.~\ref{fig1}(b)). The physics around the quadratic band touching
point can be described by an effective Hamiltonian $H_{\Gamma }=-2
k_{y}F_{x}-2k_{x}F_{y}+\left( \delta_{T}-8\right) F_{z}^{2}/4$, where two
degenerate spin states $\left\vert 1\right\rangle $ and $\left\vert
-1\right\rangle $ at $\mathbf{k}=0$ are indirectly coupled through $%
\left\vert 0\right\rangle $. Near the origin $\mathbf{k}=0$, the effective
two-level Hamiltonian (up to the second order) becomes
\begin{equation}
H_{\text{QBT}}=-2(k_{x}^{2}-k_{y}^{2})\sigma _{x}-4k_{x}k_{y}\sigma
_{y}+2(k_{x}^{2}+k_{y}^{2})I  \label{QBT}
\end{equation}%
for two touched bands. Such quadratic band touching has a winding number $-2$%
.

The Hamiltonian (\ref{QBT}) is similar as that for a quadratic band touching
in checkerboard ($C_{4}$) and Kagome ($C_{6}$) lattices, which requires
time-reversal symmetry and corresponding point group symmetry to be
topologically robust \cite{SunK2009}. In contrast, the quadratic band
touching in our model is only protected by time-reversal symmetry and robust
to the breaking of $C_{4}$ rotational symmetry because it stems from the
indirect coupling induced by extra spin degrees of freedom, as illustrated
in Fig.~\ref{fig1}(c). The green lines represent SO coupling $-\sin
(k_{y})+i\sin (k_{x})$, which contributes a winding $\pm 1$ at different
high-symmetry points. When two spins $\left\vert 1\right\rangle $ and $%
\left\vert -1\right\rangle $ are degenerate, their touching point would
naturally possess a winding number $\pm 2$ and exhibit quadratic band
touching. Nevertheless, due to the lack of point group symmetries in its
mechanism, the quadratic band touching here cannot be split into several
Dirac cones. Upon breaking time-reversal symmetry through a vector Zeeman
field $\delta _{V}$, a gapped phase with non-trivial Chern numbers for each
band appears (see Appendix~\ref{appb}).

Hereafter we use the detunings $\delta _{\pm 1}=\delta _{T}\pm \delta _{V}$
for spin states $\left\vert \pm 1\right\rangle $ from $\left\vert
0\right\rangle $, which are more relevant to realistic experimental
parameters. The single particle phase diagram for the lowest band is shown
in Fig.~\ref{fig1}(d). In the gapped phase regions, the band Chern numbers
are non-zero as long as $|\delta _{+1}|<8$ or $|\delta _{-1}|<8$. More
details about the single-particle phase diagram are presented in Appendix~%
\ref{appb}.

\section{Large Chern number 2D superfluids}

We consider two-body \textit{s}-wave attractive interaction between
Fermionic atoms. In experiments, the interaction between different spin
states can be tuned by Feshbach resonance \cite{ChinC2010,FuZ2014}. Here we
assume, without loss of generality, that the interaction $-\sum_{\bm{i}%
}U_{+1,0}n_{\bm{i},+1}n_{\bm{i},0}$ between spins $\left\vert
+1\right\rangle $ and $\left\vert 0\right\rangle $ is tuned to be dominant,
where $\bm{i}=(i_{x},i_{y})$ is the $2$D lattice-site index, $\hat{n}_{\bm{i}%
,\sigma }=\hat{c}_{\bm{i},\sigma }^{\dagger }\hat{c}_{\bm{i},\sigma }$ is
the particle number operator and $U_{+1,0}>0$ is the interaction strength.

\begin{figure}[t]
\centering\includegraphics[width=0.48\textwidth]{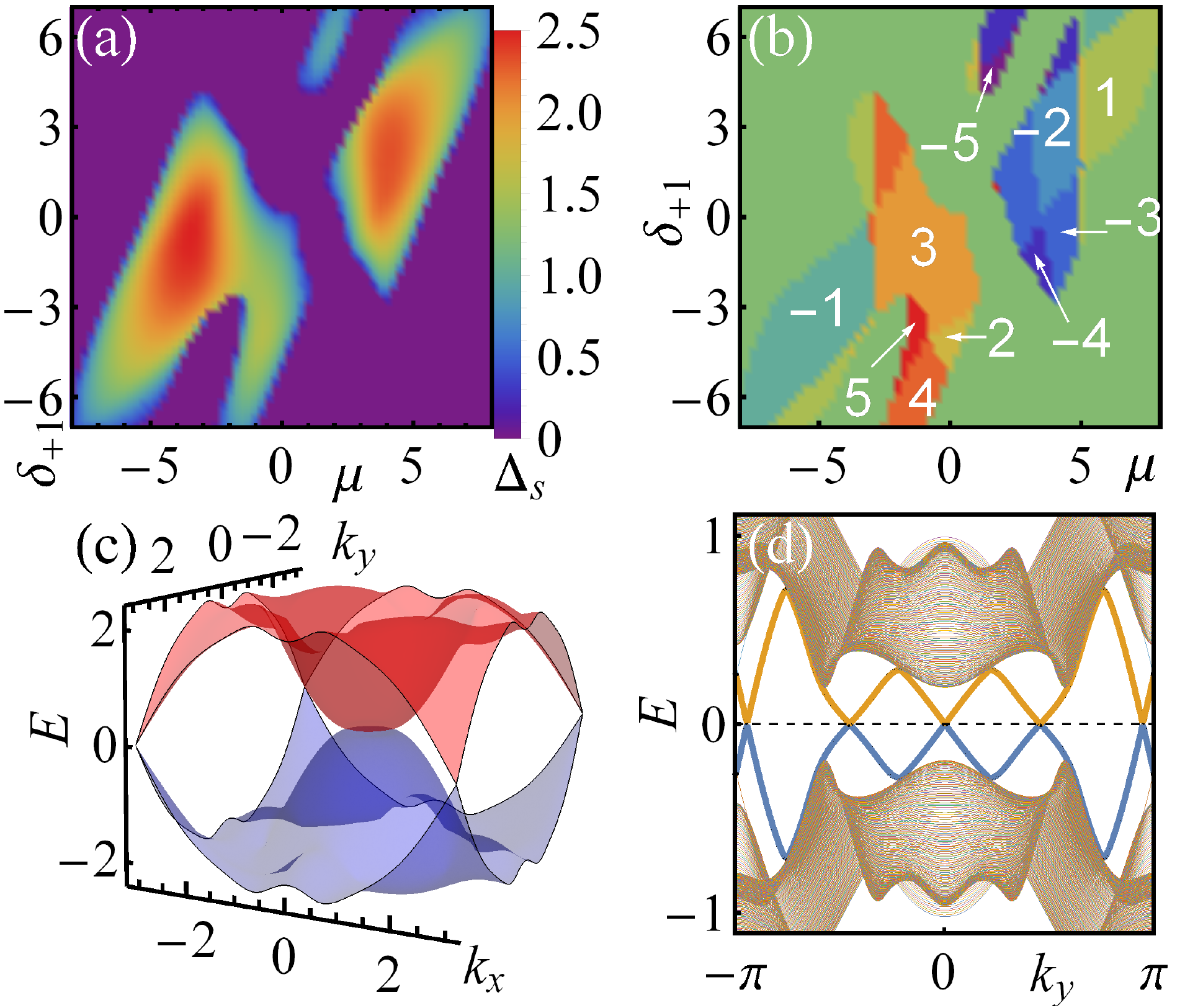}
\caption{(a,b) The phase diagrams of order parameter and Chern number with
respect to $\protect\delta _{+1}$ and $\protect\mu $ for $\protect\delta %
_{-1}=1$. The Chern number is summed over the hole branch and vanishes with
the zero order parameter. (c) Coexistence of a cubic band touching and a
Dirac point in BZ. $\protect\delta _{+1}=-2$, $\protect\delta _{-1}=1$ and $%
\protect\mu =-3$. (d) Quasiparticle spectrum with an open-boundary condition
along the $x$ direction. $\protect\delta _{+1}=-4.5$, $\protect\delta %
_{-1}=1 $ and $\protect\mu =-1.5$.}
\label{fig2}
\end{figure}

Under the mean-field approach, the Bogoliubov de-Gennes (BdG) Hamiltonian
for the 2D superfluid can be written as
\begin{eqnarray}
H_{\Delta } &=&-d_{x}F_{x}\otimes \tau _{I}-d_{y}F_{y}\otimes \tau
_{z}+d_{z}(I-2F_{z}^{2})\otimes \tau _{z}  \label{BdGH} \\
&+&(\delta _{T}F_{z}^{2}+\delta _{V}F_{z}-\mu I)\otimes \tau _{z}+\left(
\Delta _{s}F_{s}\otimes \tau _{+}+h.c.\right) ,  \notag
\end{eqnarray}%
in the Nambu basis $\Psi _{\bm{k}}=(\hat{c}_{\bm{k},+1},\hat{c}_{\bm{k},0},%
\hat{c}_{\bm{k},-1},\hat{c}_{\bm{-k},+1}^{\dagger },\hat{c}_{\bm{-k}%
,0}^{\dagger },\hat{c}_{\bm{-k},-1}^{\dagger })$, where $\mu $ is the
chemical potential, $\tau _{I}$ and $\tau _{i}$ are identity matrix and
Pauli matrices acting on Nambu space, $F_{s}=i\left(
F_{y}+\{F_{y},F_{z}/2\}\right) $, and the \textit{s}-wave superfluid order
parameter $\Delta _{s}=(U_{+1,0}/N_{0})\sum_{\bm{k}}\langle \hat{c}_{\bm{k}%
,+1}\hat{c}_{-\bm{k},0}\rangle $ with $N_{0}$ the number of atoms in spin
states $\left\vert +1\right\rangle $ and $\left\vert 0\right\rangle $.
Despite that such a pairing breaks time-reversal symmetry, the particle-hole
symmetry $\mathcal{P}=\tau _{x}\widehat{K}$ is still preserved. The order
parameter is self-consistently determined by minimizing the thermodynamical
potential \cite{Qu2013,XuY2014} and the corresponding phase diagram at $%
\delta _{-1}=1$ is plotted in Figs.~\ref{fig2}(a,b).

The coupling between different states in the above BdG Hamiltonian is
illustrated in Fig.~\ref{fig1}(c), where blue and red branches denote
particles and holes. The dashed yellow lines are couplings through order
parameter $\Delta _{s}$, which do not contribute any winding. The
highest-order band touching is then cubic, which is given by the indirect
coupling between particles and holes at spin state $\left\vert
-1\right\rangle $. Moreover, different types of band touching may appear at
different high-symmetry points at the same time. In Fig.~\ref{fig2}(c), we
show a gapless phase with both cubic band touching at $\Gamma $ and
Dirac-type linear touching at M point.

In the numerical phase diagram of the 2D superfluid (Fig.~\ref{fig2}(a,b)),
a large Chern number up to $\pm 5$ appears while the change of Chern number
may reach $6$ (from $1$ to $-5$), which is achieved through two cubic band
crossings. The effective two-level Hamiltonian around a $\Gamma $ cubic band
touching point is $\sim -(k_{y}^{3}-3k_{x}^{2}k_{y})\sigma
_{x}-(k_{x}^{3}-3k_{x}k_{y}^{2})\sigma _{y}+(k_{x}^{2}+k_{y}^{2})\sigma _{z}$%
. In the gapped region, multiple Majorana edge states emerge for the large
Chern number 2D superfluid. In Fig.~\ref{fig2}(d), we plot the band
structure for a topological superfluid with Chern number $5$ under open
boundary condition along $x$ and periodic boundary condition along $y$.
Clearly five Majorana edge states appear at each edge in the superfluid band
gap.

\begin{figure}[t]
\centering\includegraphics[width=0.48\textwidth]{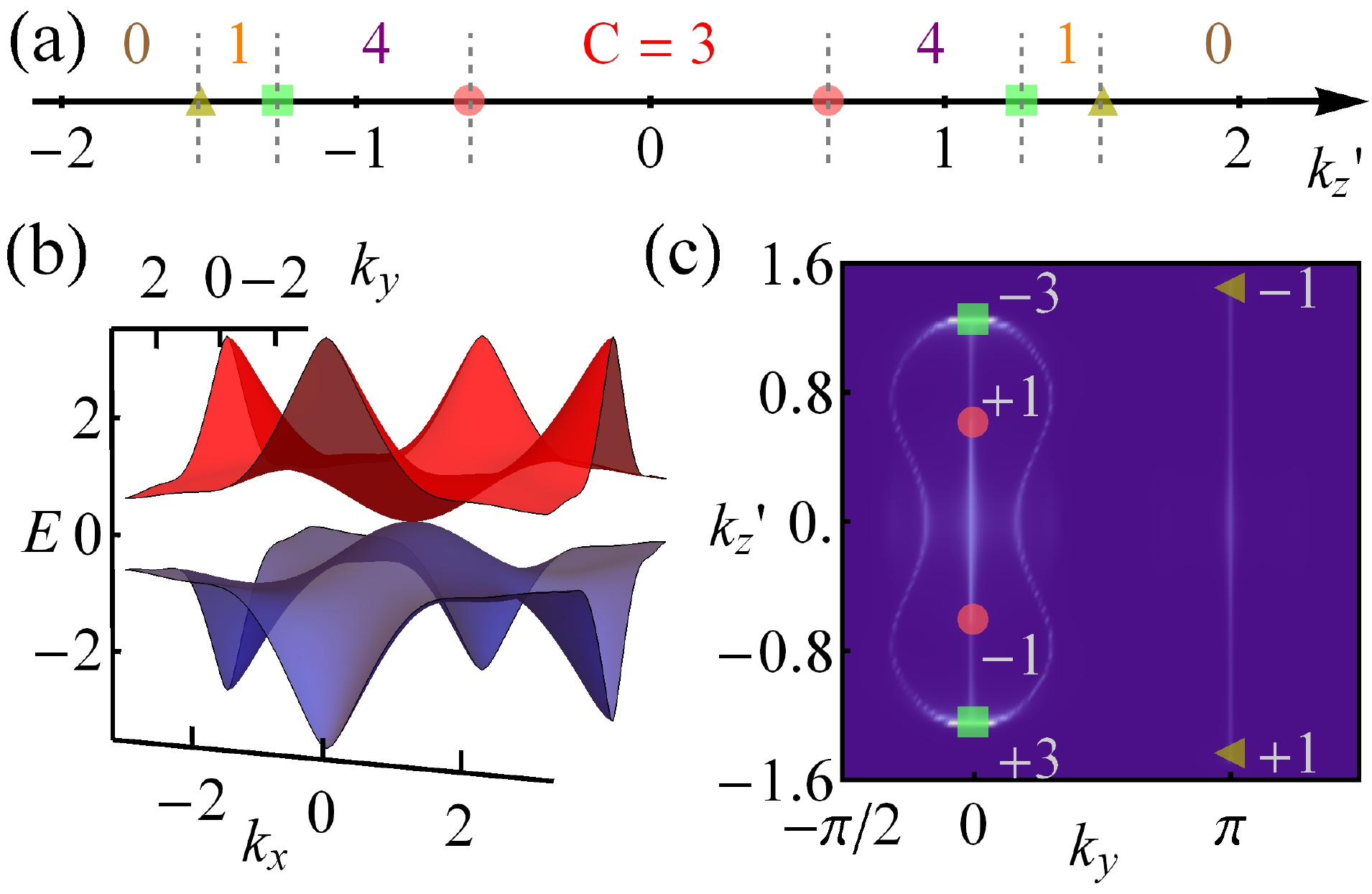}
\caption{Triple-Weyl node in a 3D topological superfluid. (a) Change of the
2D band Chern number with respect to $k_{z}^{\prime }$ due to the change of
the effective chemical potential $\protect\mu _{eff}$. Green square, red
disk, and yellow triangle denote a triple-Weyl node that locate at $\Gamma $
in the $\left( k_{x},k_{y}\right) $ plane, a Weyl point at $\Gamma$, and a
Weyl point at M, respectively. (b) Plot of the triple-Weyl node in 2D BZ at $%
k_{z}^{\prime }=1.26$, which shows cubic band dispersion along both $k_{x}$
and $k_{y}$. (c) Surface spectral densities and Fermi arcs in the $k_{y}$-$%
k_{z}^{\prime }$ surface plane with $\protect\omega =0$. $\protect\delta %
_{+1}=2$, $\protect\delta _{-1}=1$ and $\protect\mu =-1.4$.}
\label{fig3}
\end{figure}

\section{Triple-Weyl nodes in 3D superfluids}

We consider a 3D superfluid with the same 2D SO coupling and free dispersion
along the $k_{z}$ direction. Because $k_{z}$ only enters the Hamiltonian
through the kinetic energy, we can incorporate it by replacing the chemical
potential $\mu $ in the BdG Hamiltonian (\ref{BdGH}) with the effective
chemical potential $\mu _{eff}=\mu -\hbar ^{2}k_{z}^{2}/(2m)$. For
convenience, we use the coordinate $k_{z}^{\prime }=(\hbar /\sqrt{2m})k_{z}$%
. An example of the change of the 2D band topology with $k_{z}^{\prime }$ is
shown in Fig.~\ref{fig3}(a). At $k_{z}^{\prime }=0$, the 2D Chern number is
3 for the chosen chemical potential $\mu $. With increasing $k_{z}^{\prime 2}
$, $\mu _{eff}$ decreases, leading to band gap closing at different points
and the change of Chern number, as shown in Figs. \ref{fig2}(b,c). Such band
gap closing points yield linear or multi-Weyl nodes in 3D momentum space. In
total, there are three types of band touchings at different $k_{z}^{\prime }$
and they are labelled with different colored shapes in Figs. \ref{fig3}%
(a,c). Unlike multi-Weyl nodes in electronic systems, here the Weyl points
are not protected by $C_{n}$ point group symmetry, therefore we may have
triple-Weyl nodes even though our model itself exhibits only $C_{4}$,
instead of $C_{6}$ symmetry \cite{FangC2012}. Such a triple-Weyl node shows
a cubic band dispersion in the $k_{x}$-$k_{y}$ plane (Fig.~\ref{fig3}(b))
and is linear along the $k_{z}$ direction (see Appendix~\ref{appc}). By
keeping only the leading order, the two-level low-energy Hamiltonian around
the Weyl point is $\sim -(k_{y}^{3}-3k_{x}^{2}k_{y})\sigma
_{x}-(k_{x}^{3}-3k_{x}k_{y}^{2})\sigma _{y}+k_{z}^{\prime }\sigma _{z}$ up
to some constants, which is the same as that stabilized by $C_{6}$ point
group in topological semimetals \cite{FangC2012}.

In order to characterize the surface states and the triple-Weyl nodes in
Fig.~\ref{fig3}(b), we calculate and plot the spectral density function $%
A(\omega ,\mathbf{k})=\text{Im}G(i\omega ,\mathbf{k})/\pi $ at $\omega =0$
in Fig.~\ref{fig3}(c) with an open boundary condition along the $x$
direction, where $G(i\omega ,\mathbf{k})$ is the single particle Green
function. We also shift the BZ to make all surface Fermi arcs visible. The
pair of Weyl points at M (yellow triangle) gives an isolated surface arc at $%
k_{y}=\pi $. The surface arc connecting $\Gamma $ Weyl points (red disks)
overlaps with one of the three Fermi arcs connecting the $\Gamma $
triple-Weyl nodes (green square) at $k_{y}=0$, therefore the density is
slightly higher.

We remark that because the multi-Weyl nodes here do not rely on the
existence of point group symmetries $C_{n}$, they are also robust to the
breaking of $C_{4}$ symmetry, which is preserved by the system Hamiltonian.
For electronic materials with orbital degree of freedom, the highest order
for a multi-Weyl node is triple because it is stabilized through $C_{6}$
symmetry, which is the highest order allowed by classical crystalline order.
In contrast, a quadruple-Weyl or quintuple-Weyl node can be found in
principle in a spin-$3/2$ system.

\section{Further discussions}
\subsection{Experimental scheme for generating 2D SO coupling}

We briefly illustrate the experimental proposal for implementing 2D SO
coupling in Hamiltonian Equ.~\ref{ModelH}, which could be considered as a
natural generalization of the experimentally realized 2D SO coupling
for spin-1/2 atoms \cite{WuZ2016,LiuXJ2014}. The scheme is presented using $%
^{40} $K atoms, but similar setup could apply to $^{173}$Yb \cite{Jo1,Jo2}
or $^{161}$Dy \cite{Lev1} atoms, which have much less heating from Raman
lasers. More details are provided in Appendix~\ref{appd}. Our proposal focuses on lattice systems while the realization of spin-orbit coupling in free space  \cite{HuangLH2016, MengZM2016, ValdesCuriel2019} may enable the generalization and study of these distinctive band touchings in free space.

\begin{figure}[t]
\centering\includegraphics[height=1.45in,width=3.4in]{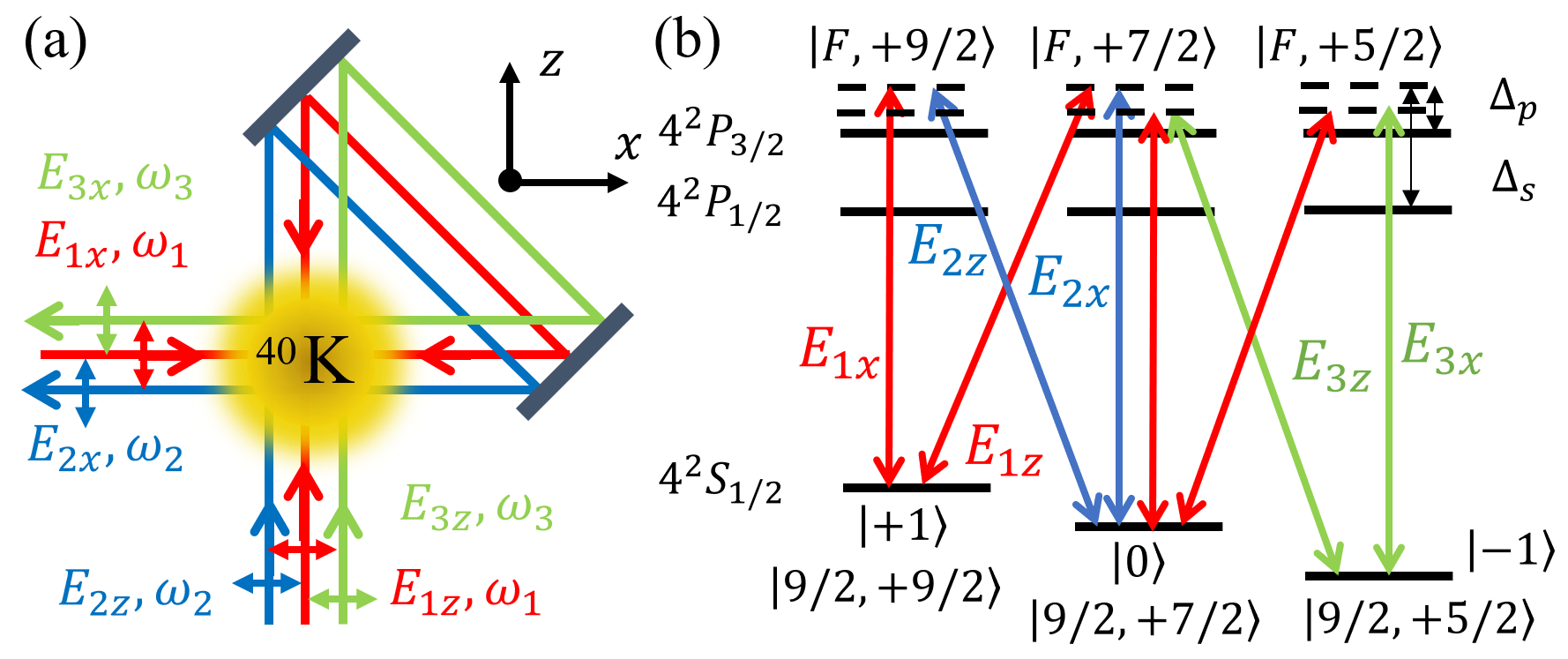}
\caption{(a) Experimental scheme for implementing 2D Rashba-type SO coupling
in fermion atoms $^{40}$K using a standing wave $\bm{E}_{1x(z)}$ and two
plane-wave $\bm{E}_{2(3)x(z)}$ laser fields. The arrows indicate the
directions of corresponding beams and each beam is reflected by two mirrors
(dark gray lines). (b) Level diagram and optical coupling in the hyperfine
structure $|F,m\rangle $ of $^{40}$K atoms.}
\label{figExp}
\end{figure}

The basic experimental setup is shown in Fig.~\ref{figExp}(a). Two
counterpropagating Raman lasers (red) form standing wave fields $\bm{E}_{1x}=%
\hat{z}E_{1x}\cos (k_{0}x)$ and $\bm{E}_{1z}=\hat{x}E_{1z}\cos (k_{0}z)$
along $x$ and $z$ directions, which also generate a spin-independent
square lattice $V(\bm{r})=V_{0x}\cos ^{2}(k_{0}x)+V_{0z}\cos ^{2}(k_{0}z)$.
As illustrated in in Fig.~\ref{figExp}(b), the red standing wave and blue plane wave $\bm{E}_{2x}=\hat{z}E_{2x}e^{ik_{0}x}$, $\bm{E}_{2z}=\hat{x}E_{2z}e^{ik_{0}z}$ (or green plane wave $\bm{E}_{3x}=\hat{z}E_{3x}e^{ik_{0}x}$, $\bm{E}_{3z}=\hat{x}E_{3z}e^{ik_{0}z}$) can induce a two-photon Raman transition between $\left\vert +1\right\rangle $ and $\left\vert 0\right\rangle$ (or $\left\vert 0\right\rangle $ and $\left\vert -1\right\rangle $). The
resulting Raman coupling can be written as $\sim
M_{x}(x,z)F_{x}+M_{y}(x,z)F_{y}$ in the spin-1 basis with $%
M_{x}(x,z)=-M_{0x}\cos (k_{0}x)\sin (k_{0}z)$ and $M_{y}(x,z)=-M_{0y}\sin
(k_{0}z)\cos (k_{0}x)$, which yield the 2D SO coupling $%
d_{x}F_{x}+d_{y}F_{y} $ in the Hamiltonian (Equ.~\ref{ModelH}) under the
tight-binding approximation (in $x$-$z$ plane). The coupling strength $M_{0x(y)}$ can be tuned
through intensity of Raman beams and optical detunings $\Delta_s$ and $\Delta_p$%
. In the lattice model, the bands between $\left\vert \pm 1\right\rangle $
and $\left\vert 0\right\rangle $ are inverted, yielding the term $%
d_{z}(I-F_{z}^{2}/2)$ in Equ.~\ref{ModelH}. The tensor and vector Zeeman fields
$\delta _{T}F_{z}^{2}/4+\delta _{V}F_{z}/2$ can be tuned by changing the two
photon Raman detunings $\delta _{\pm 1}$ between $\left\vert \pm
1\right\rangle $ and $\left\vert 0\right\rangle $. The $s$-wave pairing
interaction can be tuned through Feshbach resonance \cite%
{ChinC2010,WilliamsRA2013,FuZ2014}. To observe the topological edge state, previous experimental scheme of quenching a shaping potential in 2D square lattice can be similarly implemented \cite{GoldmanDirect2013}.

\subsection{Extension to a larger spin}

Both the physical results and proposed experimental scheme can be extended
to even higher spin systems. Here we simply list the results and leave the
details in Appendix~\ref{appe}. We consider a spin-$s$ system, where only
neighboring spins are coupled through Rashba- or Rashba-type SO coupling and
each coupling term may contribute a winding number $\pm 1$. At certain
high-symmetry point in BZ, two bands may become degenerate and a high-order
band crossing point appears with large Berry flux. Specifically, if the band
touching has a $m$-th order dispersion relation, it can possess a winding
number $m$, $m-2$, ..., $-m+2$ and $-m$, depending on the explicit form of
system Hamiltonian. Based on this argument, there are two types of quadratic
band touchings, one with $\pm 2$ winding and the other is trivial. The
low-energy Hamiltonian for the latter can be written as $k_{y}F_{x}\pm
k_{x}\{F_{y},F_{z}\}$. There are totally $2s$ SO coupling terms, therefore
the highest-order band touching should have a winding number $\pm 2s$.
Moreover, when multiple bands become degenerate at one single momentum, we
would have a topologically non-trivial and more complicated counterpart of
triply-degenerate point.

When $s$-wave attractive pairing interaction is considered, the
highest-order band crossing in the superfluid phase has the order $4s-1$
because the order parameter does not contribute any winding and the pairing
only occurs between different spin states. The extension to a multi-Weyl
node with a maximum $4s-1$ charge in a 3D superfluid is apparent. All those
exotic types of band touching points do not require any specific symmetries
like point group or inversion symmetries, but they still can be
topologically non-trivial (they do require time-reversal symmetry in certain
cases like the quadratic band touching discussed here). Therefore the large
spin systems have significant advantages over usual spin-1/2 electronic
systems on the experimental observation of novel higher-order band touchings
because the system does not have to be finely tuned to preserve certain
symmetry, for example, the equal SO coupling strengths $M_{0x}$ and $M_{0y}$
for $C_{4}$ symmetry.

\section{Conclusion}

In summary, we have studied the physics and experimental realization of
pseudospin-1 Fermi gases with 2D Rashba-type SO coupling and found many exotic
topological quantum phases, such as triply-degenerate points, quadratic and
cubic band touchings, triple-Weyl nodes, etc. Our work provides a new route
for engineering many fascinating topological quantum matters by utilizing
large spin degrees, instead of complex optical lattice geometry (see Appendix~\ref{appf} for a discussion of robustness against lattice distortions). Our results
may motivate further theoretical and experimental investigations of
interesting SO coupling effects in larger spin systems.

\section{Acknowledgments}

This work is supported by Air Force Office of Scientific Research
(FA9550-16-1-0387), National Science Foundation (PHY-1505496,PHY-1806227),
and Army Research Office (W911NF-17-1-0128).

\appendix

\section{Spin-1 Pauli matrices}

\label{appa} The spin vectors are usually defined as the finite-dimensional
irreducible representation of SU(2) which has a dimension $2s+1$ for a spin-$%
s$ systems. By convention, we denote the spin-$1/2$ spin operator as $\bm{S}%
_{1/2}=\frac{\hbar}{2}\bm{\sigma}$, where $\bm{\sigma}=(\sigma_x,\sigma_y,%
\sigma_z)$ denotes the Pauli matrices. Similarly, for a spin-1 system we
have $\bm{S}_1=\frac{\hbar}{2}\bm{F}$, where
\begin{equation}
F_x=\left(
\begin{array}{ccc}
0 & \sqrt{2} & 0 \\
\sqrt{2} & 0 & \sqrt{2} \\
0 & \sqrt{2} & 0%
\end{array}
\right), F_y=\left(
\begin{array}{ccc}
0 & -\sqrt{2}i & 0 \\
\sqrt{2}i & 0 & -\sqrt{2}i \\
0 & \sqrt{2}i & 0%
\end{array}
\right),
\end{equation}
and $F_z=\text{diag}(-2,0,2)$ is diagonal.

Note that, $F_z$ and $F_z^2$ form the Cartan subalgebra of SU(3) and thus,
any in-plane Zeeman field can be linearly decomposed up to a constant.

\section{Single-particle phase diagram}

\label{appb}

\begin{figure*}[t]
\centering\includegraphics[width=0.8\textwidth]{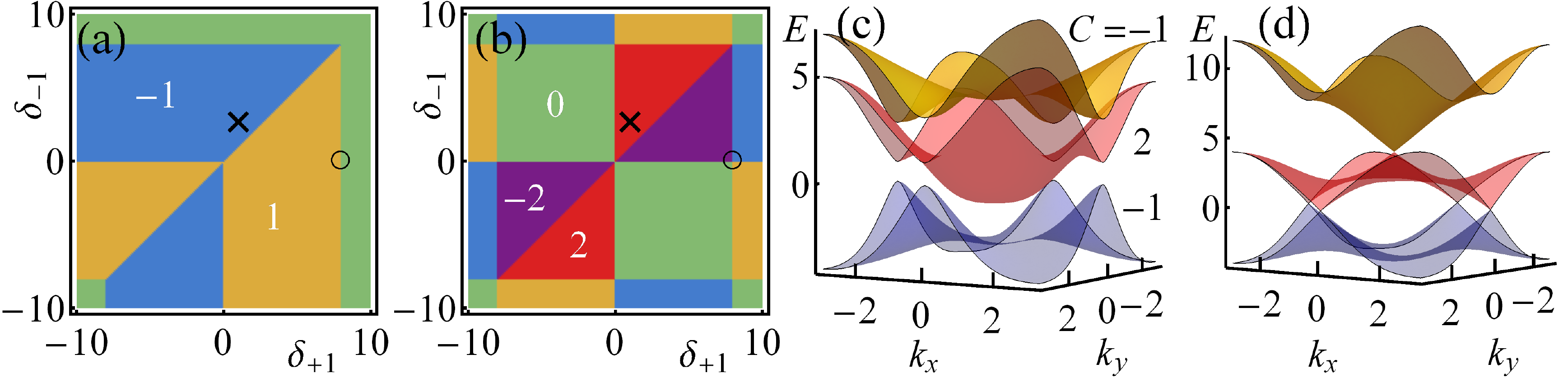}
\caption{(a) and (b): The Chern number for upper and middle bands in the
plane $(\protect\delta _{+1},\protect\delta _{-1})$, respectively. The
summation over the Chern number of all three bands equals to $0$. (c) A
typical band structure when all bands are well separated and the middle band
has a Chern number $2$. The Zeeman fields are chosen as the dark cross in
panel (a,b). (d) The single Dirac band touching between adjacent bands at
high-symmetry points in BZ. The parameters are labeled by the dark circle in
(b). The circle locates on multiple boundaries, where the middle band
crosses with both lower and upper bands.}
\label{figS1}
\end{figure*}

By tuning the detunings $\delta _{\pm 1}$, we can change the relative energy
between different bands. For a very large $\delta _{\pm 1}$, either top or
bottom band is pulled far away and the spin-1 model can be reduced to a
spin-1/2 system, which has been studied in Ref.~\cite{LiuXJ2014}. With such
observation, we expect to observe interesting topological phases in the spin-%
$1$ model when one of the detunings satisfies $0<|\delta _{\pm 1}|<8t$. In
Fig.~\ref{figS1}(a) and~\ref{figS1}(b), the phase diagrams of Chern number
for two upper bands are plotted with respect to $\delta _{\pm 1}$. Note that
the phase diagram for the lowest band was presented in the main text.

Since the upper and lower bands are only coupled to the middle band, similar
as the spin-$1/2$ case, they should have $C=\pm 1$ when $0<|\delta _{\pm
1}|<8t$. If the Chern numbers of those two bands have opposite sign, the
middle band must be trivial. Otherwise, the middle band has a large Chern
number $C=\pm 2$ in the opposite way, as illustrated in Fig.~\ref{figS1}(c).
Such a combination makes the phase diagram of the middle band much richer.
If we consider only the topological phase transition points from the spin-$%
1/2$ case, i.e., $\delta _{\pm 1}=0,\pm 8t$, the $(\delta _{+1},\delta
_{-1}) $ plane is divided in to $16$ square (rectangle) regions with
different Chern number $\pm 2$, $\pm 1$ and $0$. In most cases, the
transition is characterized by emergence of Dirac cones at high-symmetry
points in BZ. An interesting example in which the middle band touches both
lower and upper bands is shown in Fig.~\ref{figS1}(d).

Each Dirac cone carries a Berry flux $\pm \pi $, which changes the Chern
number by $\pm 1$. In this sense, when Chern number is changed by $\pm 2$ ($%
\pm 2$ to $0$ or $\pm 1$ to $\mp 1$), a pair of Dirac cones must appear.
Note that, unlike a spin-1/2 system, the Dirac cone here is not protected by
the time-reversal symmetry, although its low-energy Hamiltonian does exhibit
such a symmetry. For the phase transition from Chern number $\pm 2$ to $\mp
2 $, a possible mechanism is that four Dirac cones appear with each one
contributing a change of $1$. However, this is not the case here. In our
model, four high-symmetry points fall into two groups, which are mutually
exclusive for weak Zeeman fields. Therefore we cannot expect to have four
same type of band touching points between two bands, and higher-order band
touching points with quantized non-zero (and $>\pi $) Berry flux must appear
to have the large change ($\pm $4) of the Chern number. Such quadratic band
touching and associated band structure have been studied explicitly in the
main text.

Similarly, with increasing $\delta _{T}$, the lower band is pulled away from
the triply-degenerate point and upper two bands are degenerate with
quadratic band dispersion. However, band Chern number vanishes after the
gapless points are open because the upper two bands have quadratic band
touchings at all $\Gamma $, M and two X points with wind number $-2$, $-2$
and $2\ast 2$, whose summation is zero.

\section{(Triple-)Weyl points and edge states}

\label{appc}

\begin{figure*}[t]
\centering\includegraphics[width=0.7\textwidth]{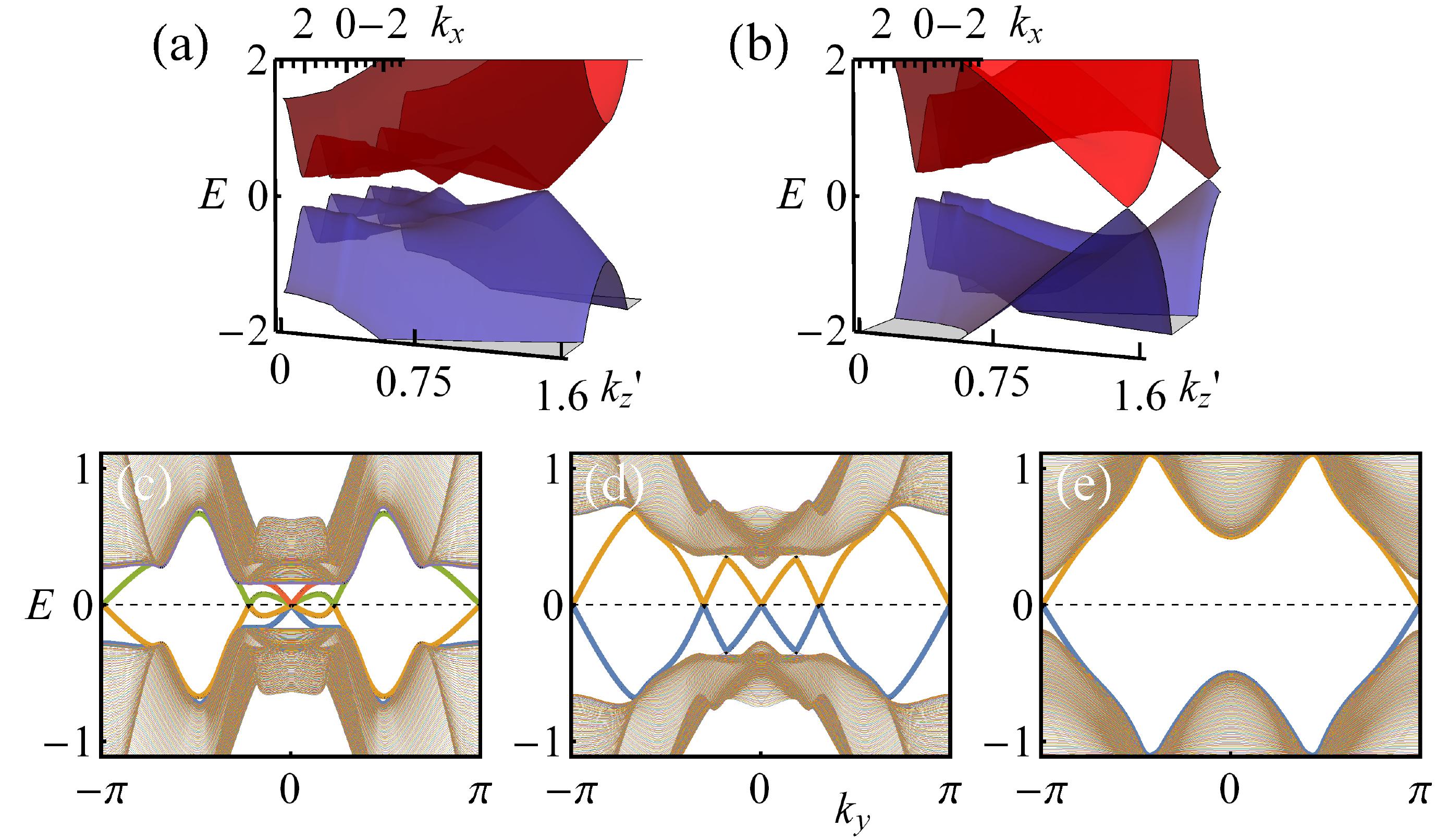}
\caption{(a) Band structure in the $k_{y}=0$ plane. The triple-Weyl node
locates on the right side and the dispersion is linear along the $%
k_{z}^{\prime }$ direction. On the left-hand side, there is a Weyl point
with linear dispersions along all directions. (b) Band structure in the $%
k_{y}=\protect\pi $ plane with a M Weyl point. (c-e) Edge states along $%
k_{y} $ with fixed $k_{z}^{\prime }=0.45$, $k_{z}=0.77^{\prime }$ and $%
k_{z}^{\prime }=1.45$ respectively. The other parameters are the same as
those in Fig.~3.}
\label{figS2}
\end{figure*}

In Fig.~\ref{figS2}(a), the triple-Weyl point shown in Fig.~3(b) is also
plotted in the $k_{z}^{\prime }$-$k_{x}$ plane (the right touching point),
which shows a linear band dispersion along $k_{z}^{\prime }$. The band
touching on the left is a Weyl point, indicated by the red disks in
Figs.~3(a,c). Another Weyl point at larger $|k_{z}|$ (yellow triangles in
Figs.~3(a,c)) appears at M point $k_{x}=k_{y}=\pi $ as shown in Fig.~\ref%
{figS2}(b).

The corresponding edge states for a few 2D band structures with fixed $%
k_{z}^{\prime }$ are shown in Figs.~\ref{figS2}(c-e). When $k_{z}^{\prime }$
lies in between two $\Gamma $ Weyl points, there are three edge modes Fig.~%
\ref{figS2}(c), agreeing with the bulk Chern number 3 (Fig.~3(a)). However,
one of these edge modes crosses zero energy twice, leading to four surface
arcs observed in Fig.~3(c) and large surface state density at $k_{y}=0$.
Across the $\Gamma $ Weyl point, the band gap at $k_{y}=0$ for the twisted
edge state is opened, and the number of edge states becomes four, agreeing
with the bulk Chern number 4 in Fig.~3(a). When $k_{z}^{\prime }$ further
increases and we cross the $\Gamma $ tripe-Weyl node, only 1 edge states at $%
k_{y}=\pi $ is left as shown in Fig.~\ref{figS2}(e), agreeing with the bulk
Chern number 1.

\section{Effective Hamiltonian and tight-binding model}

\label{appd}

\label{app1} As illustrated in Fig.~4(a), two beams (red lines) are incident
from both $x$ and $z$ directions and reflected by two mirrors to form two
standing waves $\bm{E}_{1x}=\hat{z}E_{1x}e^{i(\varphi _{1x}+\varphi
_{1z}+\varphi _{L})/2}\cos (k_{0}x+\alpha )$ and $\bm{E}_{1z}=\hat{x}%
E_{1z}e^{i(\varphi _{1x}+\varphi _{1z}+\varphi _{L})/2}\cos (k_{0}z+\beta )$%
, where $E_{1x(z)}$ is field strength, $\varphi _{1x(z)}$ is the initial
phase, $\varphi _{L}=k_{0}K$ is the phase picked up from optical path $K$
and $\alpha (\beta )=(\varphi _{1x(z)}-\varphi _{1z(x)}-\varphi _{L})/2$.
Another two beams (blue and green lines) are incident along $z$ direction
to form plane-waves $\bm{E}_{2(3)z}=\hat{x}E_{2(3)z}e^{i(k_{0}z+\varphi
_{2(3)})}$ and $\bm{E}_{2(3)x}=\hat{z}E_{2(3)x}e^{i(-k_{0}x+\varphi
_{2(3)}+\varphi _{L}-\delta \varphi _{L2(3)})}$ with the initial phases $%
\varphi _{2(3)}$ and relative phases $\delta \varphi _{L2(3)}=(\omega
_{1}-\omega _{2(3)})K/c$.

The level diagram and optical couplings are illustrated in Fig.~4(b).
Although both transition lines $D_{2}({}^{2}S_{1/2}\rightarrow {}^{2}P_{3/2})
$ and $D_{1}({}^{2}S_{1/2}\rightarrow {}^{2}P_{1/2})$ contribute to the
coupling, the latter is negligible due to large detuning $\Delta$. As a
result, we mainly consider the contribution from $D_{2}$ transitions. The
optical dipole potential are summed over all possible transitions
\begin{equation}
V_{m_{\sigma },T}(x)=\frac{|\Omega _{1x,m_{\sigma },T}|^{2}}{\Delta _{p}}%
,V_{m_{\sigma },T}(z)=\frac{|\Omega _{1z,m_{\sigma },T}|^{2}}{\Delta _{p}},
\end{equation}%
where $m_{\sigma }=+9/2,+7/2,+5/2$ denotes the value of $m_{F}$ for each
spin component (corresponding to $+1$, $0$ and $-1$, respectively) and $%
T=\sigma ^{\pm },\pi $ represents three different transitions. The effective
Rabi frequency $\Omega _{1x(z),m_{\sigma },T}$ is defined through
\begin{eqnarray}
&&\Omega _{1x,m_{\sigma },T}=\sqrt{\sum_{F}|\Omega _{x,F,m_{\sigma },T}|^{2}}%
, \\
&&\Omega _{1y,m_{\sigma },T}=\sqrt{\sum_{F}|\Omega _{z,F,m_{\sigma },T}|^{2}}%
,F=\frac{11}{2},\frac{9}{2},...,m_{\sigma },  \notag
\end{eqnarray}%
with
\begin{eqnarray}
&&|\Omega _{x(z),F,m_{\sigma },\sigma ^{+}}| =|\mu _{m_{\sigma },F,\sigma
^{+}}||E_{1x(z)}|,  \notag \\
&&|\mu _{m_{\sigma },F,\sigma ^{+}}|=\langle \frac{9}{2},m_{\sigma }|\bm{r}%
\cdot \epsilon _{1x(z)}|F,m_{\sigma }+1\rangle ,  \notag \\
&&|\Omega _{x(z),F,m_{\sigma },\pi }| =|\mu _{m_{\sigma },F,\pi
}||E_{1x(z)}|,  \notag \\
&&|\mu _{m_{\sigma },F,\pi }|=\langle \frac{9}{2},m_{\sigma }|\bm{r}\cdot
\epsilon _{1x(z)}|F,m_{\sigma }\rangle ,  \notag \\
&&|\Omega _{x(z),F,m_{\sigma },\sigma ^{-}}| =|\mu _{m_{\sigma },F,\sigma
^{-}}||E_{1x(z)}|,  \notag \\
&&|\mu _{m_{\sigma },F,\sigma ^{-}}|=\langle \frac{9}{2},m_{\sigma }|\bm{r}%
\cdot \epsilon _{1x(z)}|F,m_{\sigma }-1\rangle,  \notag
\end{eqnarray}%
where $\epsilon _{1x(z)}$ are the polarization vectors of lasers. For a $\pi
$ transition, we have
\begin{eqnarray}
&&V_{+1,\pi ,x(z)}:V_{0,\pi ,x(z)}:V_{-1,\pi ,x(z)} \\
&=&\sum_{F}|\mu _{9/2,F,\pi }|^{2}:\sum_{F}|\mu _{7/2,F,\pi
}|^{2}:\sum_{F}|\mu _{5/2,F,\pi }|^{2},  \notag
\end{eqnarray}%
which is
\begin{eqnarray}
&&V_{+1,\pi ,x(z)}:V_{0,\pi ,x(z)}:V_{-1,\pi ,x(z)}  \notag \\
&=&1215+3240:2187+1960+308:2916+1000+539  \notag \\
&=&1:1:1  \notag
\end{eqnarray}%
for the experimental data of ${}^{40}$K. Similarly, we can calculate those
coefficients for $\sigma ^{\pm }$ transitions
\begin{eqnarray}
&&V_{+1,\sigma ^{\pm },x(z)}:V_{0,\sigma ^{\pm },x(z)}:V_{-1,\sigma ^{\pm
},x(z)}  \notag \\
&=&13365+243+1440+2772:  \notag \\
&&10935+1440+729+2560+2156:  \notag \\
&&8748+2560+77+1458+3360+1617  \notag \\
&=&1:1:1.  \notag
\end{eqnarray}%
One can also verify that this still holds true even when we take $D_{1}$
line into account. Therefore this lattice potential is indeed
spin-independent and can be written as
\begin{equation}
V(\bm{r})=V(x)+V(z)=V_{0x}\cos ^{2}(k_{0}x+\alpha )+V_{0z}\cos
^{2}(k_{0}z+\beta ).
\end{equation}

As shown in Fig.~4(b), each plane-wave induces two Raman couplings. The four
coupling strengths are
\begin{eqnarray}
M_{1x,2z} &=& \sum_{F}\frac{\Omega _{1x,F,9/2}^{\ast }\Omega _{2z,F,7/2}}{%
\Delta _{p}},  \notag \\
M_{1z,2x} &=& \sum_{F}\frac{\Omega _{1z,F,9/2}^{\ast }\Omega _{2x,F,7/2}}{%
\Delta _{p}},  \notag \\
M_{1x,3z} &=& \sum_{F}\frac{\Omega _{1x,F,7/2}^{\ast }\Omega _{3z,F,5/2}}{%
\Delta _{p}},  \notag \\
M_{1z,3x} &=& \sum_{F}\frac{\Omega _{1z,F,7/2}^{\ast }\Omega _{3x,F,5/2}}{%
\Delta _{p}},  \notag
\end{eqnarray}%
where
\begin{eqnarray}
\Omega _{ix,F,\sigma _{m}} &=&\langle \frac{9}{2},m_{\sigma }|\hat{x}\cdot
\epsilon _{ix}|F,m_{\sigma }\rangle E_{ix},i=1,2,3;  \notag \\
\Omega _{1z,F,\sigma _{m}} &=&\langle \frac{9}{2},m_{\sigma }|\hat{z}\cdot
\epsilon _{1z}|F,m_{\sigma }-1\rangle E_{1z},  \notag \\
\Omega _{iz,F,\sigma_{m}}&=&\langle \frac{9}{2},m_{\sigma }|\hat{z}\cdot
\epsilon _{iz}|F,m_{\sigma}+1\rangle E_{iz},i=2,3;  \notag
\end{eqnarray}%
After inserting the effective Rabi frequency, we obtain
\begin{eqnarray}
M_{2(3)z,1x} &=&M_{0,2(3)x}  \notag \\
&&\times \cos (k_{0}x+\alpha)e^{-i(k_{0}z+\beta)}
e^{i(\varphi_{2(3)}-\varphi_{1z})},  \notag \\
M_{2(3)x,1z} &=&M_{0,2(3)z}  \notag \\
&&\times \cos (k_{0}z+\beta )e^{i(k_{0}x+\alpha )}e^{i(\delta \varphi
_{L2(3)}+\varphi _{2(3)}-\varphi _{1z})}.  \notag
\end{eqnarray}%
Note that terms proportional to $\cos (k_{0}x)\cos (k_{0}z)$ are
antisymmetric to each lattice site in both $x$ and $z$ directions and thus
can be neglected for low-band physics \cite{LiuXJ2014}. The resulting
coupling strengths are
\begin{eqnarray}
\mathcal{M}_{2(3)x} &=&-M_{2(3)x}+M_{2(3)y}\cos \delta \varphi _{L2(3)}, \\
\mathcal{M}_{2(3)y} &=&M_{2(3)y}\sin \delta \varphi _{L2(3)},
\end{eqnarray}%
with $M_{2(3)x}=M_{0,2(3)x}\cos (k_{0}x+\alpha )\sin (k_{0}z+\beta )$ and $%
M_{2(3)y}=M_{0,2(3)y}\cos (k_{0}z+\beta )\sin(k_{0}x+\alpha )$. In the
following, we assume that the strengths of incident beams are tuned such
that $\mathcal{M}_{2x(y)}=\mathcal{M}_{3x(y)}=\mathcal{M}_{x(y)}$. Now, the
total effective Hamiltonian in 2D can be written as
\begin{equation}
H=\frac{\bm{p}^{2}}{2m}+V(\bm{r})+\mathcal{M}_{x}F_{x}+\mathcal{M}%
_{y}F_{y}+\delta _{T}F_{z}^{2}+\delta _{V}F_{z}.
\end{equation}

If $\delta \varphi _{L}=n\pi ,n\in \mathbb{Z}$, the SO coupling becomes 1D. Here, we
set $\delta \varphi _{L}=\pi /2$ and $\alpha ,\beta =2n\pi $ such that the
coupling terms become
\begin{eqnarray}
M_{x}(x,z) &=&-M_{0x}\cos (k_{0}x)\sin (k_{0}z), \\
M_{y}(x,z) &=&-M_{0y}\cos (k_{0}z)\sin (k_{0}x).
\end{eqnarray}%
As we only consider the lowest $s$-orbital $\phi _{s,\sigma }$ ($\sigma
=+1,0,-1$) and nearest-neighbor hopping, the tight-binding Hamiltonian is
\begin{eqnarray}
H_{\text{TI}}&=&-\sum_{\langle \bm{i},\bm{j}\rangle ,\sigma }t^{\bm{ij}}\hat{%
c}_{\bm{i},\sigma }^{\dagger }\hat{c}_{\bm{j},\sigma } \\
&+&\sum_{\langle \bm{i},\bm{j}\rangle }\Big(t_{\text{so},+}^{\bm{ij}}\hat{c}%
_{\bm{i},+1}^{\dagger }\hat{c}_{\bm{j},0}+h.c.+t_{\text{so},-}^{\bm{ij}}\hat{%
c}_{\bm{i},0}^{\dagger }\hat{c}_{\bm{j},-1}+h.c.\Big)  \notag \\
&+&\delta _{T}\sum_{\bm{i}}\left( \hat{n}_{\bm{i},+1}+\hat{n}_{\bm{i}%
,-1}\right) +\delta _{V}\sum_{\bm{i}}\left( \hat{n}_{\bm{i},+1}-\hat{n}_{%
\bm{i},-1}\right) ,  \notag
\end{eqnarray}%
where hopping strengths can be expressed as overlap integrals
\begin{equation}
t^{\bm{ij}}=\int d^{2}\bm{r}\phi _{s,\sigma }^{\bm{i}}(\bm{r})\left[ \frac{%
\bm{p}^{2}}{2m}+V(\bm{r})\right] \phi _{s,\sigma }^{\bm{i}}(\bm{r}),
\end{equation}%
and
\begin{eqnarray}
t_{\text{so},+}^{\bm{ij}} &=&\int d^{2}\bm{r}\phi _{s,+1}^{\bm{i}}(\bm{r})%
\left[ M_{x}(x,z)F_{x}+M_{y}(x,z)F_{y}\right] \phi _{s,0}^{\bm{i}}(\bm{r}),
\notag \\
t_{\text{so},-}^{\bm{ij}} &=&\int d^{2}\bm{r}\phi _{s,0}^{\bm{i}}(\bm{r})%
\left[ M_{x}(x,z)F_{x}+M_{y}(x,z)F_{y}\right] \phi _{s,-1}^{\bm{i}}(\bm{r}).
\notag
\end{eqnarray}%
While the usual nearest-neighbor hopping is obviously the same at different
sites in all directions, the spin-flip process is more subtle. Even though
it has been generalized to spin-$1$ system, the reasoning in Ref.~\cite%
{WuZ2016} is still valid and thus, the Raman potential hopping is staggered
as
\begin{equation}
t_{\text{so}}^{i_{x},i_{x}\pm 1}=\pm (-1)^{i_{x}+i_{z}}t_{\text{so}},t_{%
\text{so}}^{i_{z},i_{z}\pm 1}=\pm i(-1)^{i_{x}+i_{z}}t_{\text{so}}.
\end{equation}%
Upon applying the transformations $\hat{c}_{\bm{i},0}\rightarrow e^{i\pi
(x_{i}+z_{i})}\hat{c}_{\bm{i},0}$ and $\hat{c}_{\bm{i},-1}\rightarrow
e^{i2\pi (x_{i}+z_{i})}\hat{c}_{\bm{i},-1}$ and a Fourier transformation, we
obtain our model Hamiltonian in momentum-space up to some constants (where we have also defined lattice constant $a=\frac{\pi}{k_0}$ as the unit of length). We remark that such an unitary transformation does not affect the form of
interatomic interaction.

\section{Extension to larger spins}

\label{appe}

In the last subsection in the main text, we generalize our results on
high-order band touching to a genuine large spin system and compare them
with those in electronic systems enriched by special point group symmetry $%
C_{n}$. This section provides some detailed discussions.

Considering a spin-$s$ system, our model Equ.~1 can be extended to $%
H_{0}=-\sum_{\bm{k},l\neq j}h_{\bm{k},l,j}\hat{c}_{\bm{k},l}^{\dagger }\hat{c%
}_{\bm{k},j}$ where the spin indices $-s\leq l,j\leq s$ enumerate each spin
component. The matrix element of $h_{\bm{k},l,j}$ can be written as $%
h_{l,l}=\delta _{l}+t_{l}(\cos (k_{x})+\cos (k_{y}))$, $%
h_{l,l+1}=t_{l,l+1,y}\sin (k_{y})+it_{l,l+1,x}\sin (k_{x})$ and $%
h_{l,l+1}=h_{l+1,l}^{\ast }$. In the following, we drop the subscript $\bm{k}
$ for convenience and assume all coupling constants are real. The coupling
terms $t_{l,l+1,x}$ and $t_{l,l+1,y}$ are generic since we impose no special
symmetry. A band touching point appears at the two X ($\Gamma $ and M)
points for $\delta _{l}=\delta _{j}$ ($\delta _{l}+2t_{l}=\delta _{j}+2t_{j}$
and $\delta _{l}-2t_{l}=\delta _{j}-2t_{j}$). In our previous discussion, $%
t_{l}=t_{j}$, therefore the band touchings at $\Gamma $ and M always
accompany with each other.

We start from the case where degeneracy only happens between two spin
components $l$ and $j$ ($j>l$). Expanding the Hamiltonian around a band
touching point at one of the four high-symmetry points $\bm{K_n}$, one has
the following two-level effective Hamiltonian
\begin{equation}
H(\bm{K_n}+\bm{q})=g(\bm{q})\sigma _{+}+g^{\ast }(\bm{q})\sigma _{-}+\delta
\sigma _{z},  \label{general}
\end{equation}%
where $\sigma _{\pm }=\sigma _{x}\pm i\sigma _{y}$ and $q=|\bm{q}|$ is
assumed to be small. Any other diagonal term is neglected and $\delta $ is
introduced for later convenience. So far we discuss $\delta =0$ for the
single particle case. The coupling term would be in the form $g(\bm{q})=\Pi
_{l\leq k<j}(a_{k}qe^{i\theta _{k}}-b_{k})$ in general and the constants $%
b_{k}$ come from expanding trigonometric function around $\bm{K_n}$. Notice
that this gives, up to some constants, the Cartesian products of three real
irreducible representations of point group $C_{n}$, therefore it is not
surprising to see the low-energy Hamiltonian of high-order band touching
point in our model shares the same form as those stabilized by $C_{n}$
rotational symmetry \cite{SunK2009,FangC2012}. When $a_{k}\neq 0$ and each $%
\theta _{k}$ can be well-defined, this high-order band touching point has a
multiplicity $m=j-l$. Note that, generally, this multiplicity is not equal
to the Berry flux of such a crossing point. In fact, the winding number of
this touching point is $\sum_{k}\text{sign}(\theta _{k})$. If $\text{sign}%
(\theta _{k})$ can be tuned (i.e., the relative sign between $t_{k,k+1,x}$
and $t_{k,k+1,y}$), we are able to engineer a $m$-order band touching point
with possible Berry flux $m\pi $, $(m-2)\pi $, ..., $(-m+2)\pi $, $-m\pi $.
For example, for the coupling $k_{x}F_{y}+k_{y}F_{x}$ and in the region $%
\text{sign}(\delta _{+1})\text{sign}(\delta _{-1})>0$ and $-8<\delta _{\pm
1}<8$, we have a large Chern number phase with the phase transition
characterized by quadratic band touching carrying a $-2\pi $ Berry flux.
However, if the model is modified to $k_{y}F_{x}\pm k_{x}\{F_{y},F_{z}\}$
(time-reversal symmetry has been broken), the band touching point still
shows a quadratic dispersion, but becomes topologically trivial.

Now, assume we have multiple degeneracy among spin components $-s\leq
s_{1}<s_{2}<...<s_{n}\leq s$ when $\delta _{s_{k}}=\delta
_{c,s_{k}},s_{k}\in S=\{s_{1},s_{2},...,s_{n}\}$ at a certain $\bm{K_n}$ in
BZ. The whole phase diagram for a given band would have dimension $2s$. We
first consider one ordered pair $s_{i}$ and $s_{j}$ with $s_{j}>s_{i},
s_{i},s_{j}\in S$. Following what we have discussed above, this defines a
gapless ($2s-1$ dimensional) subspace $R_{s_{i},s_{j}}$ with a $s_{j}-s_{i}$
order band touching point carrying some winding number $w_{s_{i},s_{j}}$ if $%
\delta _{s_{k}}$ is slightly deviated from $\delta _{c,s_{k}}$ for any $%
s_{k}\in S$ and $s_{k}\neq s_{i},s_{j}$. The multiply degenerate point
exists in the subspace expressed as $\cup _{(s_{i},s_{j})}R_{(s_{i},s_{j})}$
when $\delta _{s_{k}}$ approaches $\delta _{c,s_{k}}$ from all allowed
directions and has a winding number $w_{s_{0},s_{n}}$, which is equal to $%
\sum_{1\leq i<n}w_{s_{i},s_{(i+1)}}$. Thus, one would obtain a topologically
nontrivial multiply degenerate point as a generalization of
triply-degenerate point in a lager spin system.

So far we have seen that the high-order band touching point conceived in our
system can be ascribed to degeneracy and indirect coupling between
degenerate spin components. Thus, it would be natural to have even higher
order band crossing when pairing is introduced. The pairing order parameter $%
\Delta _{i,j}$ between spin component $i$ and $j$ enters the BdG Hamiltonian
as complex numbers and opens superconducting gaps. We would consequently see
many gapped topological superfluids with large Chern numbers and high-order
band touching points that serve as topological phase transition points.
However, $\Delta _{i,j}$ is a constant and contributes neither multiplicity
nor winding to a band touching point. In previous discussions, we only
consider pairing between $|0\rangle $ and $|1\rangle $ to have a cubic band
touching.

When the third spatial dimension is included, it appears in the diagonal
terms as free single particle kinetic energy. After expanding system
Hamiltonian around a band touching at $\bm{K_n}$ and some $k_{z0}^{\prime }$
we have $\delta =k_{z}^{\prime }$ in Eq.~\ref{general}. Now, Eq.~\ref%
{general} is a direct generalization of chiral Weyl fermions in free space.
Such a multi-Weyl node may carry a large charge and is spatially
anisotropic, i.e., it is linear along $k_{z}$ but shows high-order
dispersion along $k_{x}$ and $k_{y}$. We expect exotic chiral magnetic
effects due to the non-linear band structure of multi-Weyl nodes, as
compared to traditionally defined Weyl fermions \cite{HuangZM2017}.

\section{Robustness of band touching against distorted lattice} \label{appf}
In the main text, we consider the SO coupled Fermi gases on a regular square lattice with $C_4$ symmetry. While we have provided physical insights into the symmetry protection using the coupling scheme depicted in Fig.~\ref{fig1}(c), we offer a numerical verification in this appendix. 

\begin{figure}[h]
\centering\includegraphics[width=0.48\textwidth]{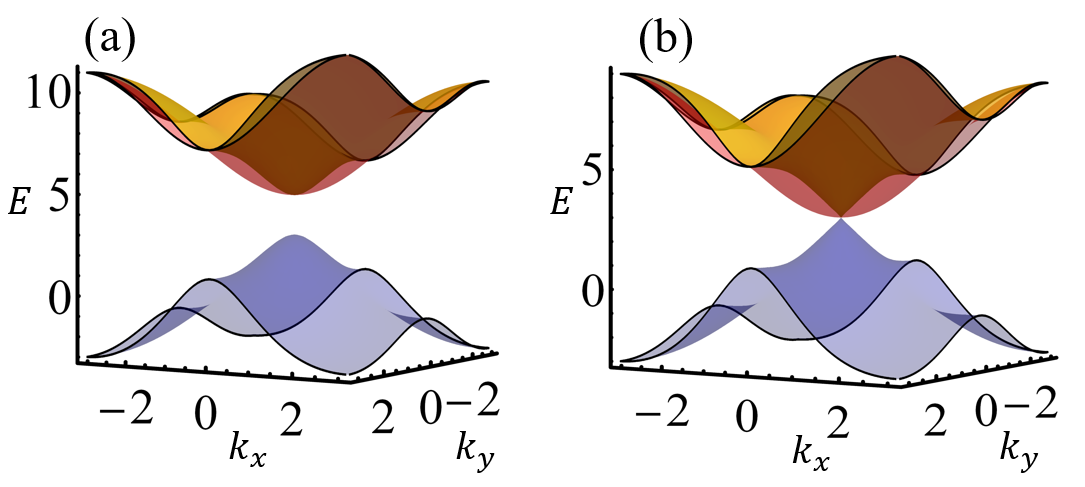}
\caption{(a) Similar to Fig.~\ref{fig1}(a) but plotted with symmetry breaking terms such that $t_x=1$, $t_y=0.5$, $t_{SO,x}=1$ and $t_{SO,y}=0.5$. (b) With such an anisotropy, the 2D triply degenerate point now occurs at $\delta_T=6$.}
\label{figR1}
\end{figure}

To show that the rotational symmetry does not affect the band touchings studied in Sec.~\ref{sec_sp}, here, we distort lattice potential to be anisotropic. Consequently, both the bare hopping and SO coupling along $x$ and $y$ are no longer the same. We compare in Fig.~\ref{fig1}(a) and Fig.~\ref{figR1}(a) how such an anisotropy affects the single-particle band structures. The quadratic band touchings at the corners and the sides of the Brillouin zone survive while it seems apparently that the 2D triply-degenerate point vanishes. In fact, the triply-degenerate point is merely shifted along the time-reversal symmetry line in parameter space and now locates at $\delta_T=6$.

\end{document}